\documentclass{elsart}

\usepackage{graphics}
\usepackage{graphicx}
\usepackage{amsmath,amssymb,epsfig,picinpar,graphics,float,amssymb,verbatim}

\newcommand{\nsum}[1]{\langle\,{#1}\,\rangle}
\newcommand{\ma}[1]{\mathcal{#1}}

\begin{document}
\begin{frontmatter}

\title{Effects of microscopic disorder on the collective dynamics of globally coupled maps}

\author{Silvia De Monte$^{a,b}$, Francesco d'Ovidio$^{c}$, Hugues
 Chat{\'e}$^{b}$, Erik Mosekilde$^a$} 

\address{$^a$ Dept. of Physics, 
The~Technical~University~of~Denmark, DK 2800 Lyngby, Denmark\\
$^b$ CEA -- Service de Physique de l'Etat Condens\'e, 
Centre d'Etudes de Saclay, 91191 Gif-sur-Yvette, France\\
$^c$  IMEDEA, CSIC University of Balearic Islands, E~07071~Palma de Mallorca, Spain
}
\date{\today}

\begin{abstract}
This paper studies the effect of independent additive noise on the
synchronous dynamics of large populations of globally coupled maps.
Our analysis is complementary to the approach taken by Teramae and Kuramoto
\protect\cite{teramae01}
who pointed out the anomalous scaling properties preceding 
the loss of coherence.
We focus on the macroscopic dynamics that remains deterministic at any noise
level and differs from the microscopic one.
Using properly defined order parameters, 
an analytical approach is proposed for
describing the collective dynamics in terms of 
an approximate low-dimensional system. The  
systematic derivation of the macroscopic equations 
provides a link between the microscopic features of the population 
(single-element dynamics and noise
distribution) and the properties of the emergent behaviour. 
The macroscopic bifurcations induced by noise are compared to those
originating from parameter mismatches within the population.
 
\end{abstract}


\end{frontmatter}

\begin{section}{Introduction}

A large number of dynamical phenomena are accurately described by
deterministic equations.
In many cases, however, the deterministic structure of the dynamics is
altered by the presence of degrees of freedom others than those included in
the equations. Such degrees of freedom cannot be explicitly formulated in
the model because their features and actual influence on the observable
time evolution are not known. It is commonly assumed that their effect on the
deterministic dynamics can be accounted for by additive or multiplicative
stochastic terms.

Besides studies of the effects of noise on a single low-dimensional
dynamical system, it is only recently that significant attention 
has been paid to the investigation of
the influence of stochastic perturbations 
on large ensembles of interacting dynamical elements
\cite{anishchenko02,lindner04}. Here we report on such investigations
on large populations of globally coupled nonlinear maps.

Such an interaction scheme, where each population element
only feels the average state of the others, has been widely used to
model a range of physical, chemical and biological systems,
such as arrays of semiconducting elements 
(Josephson junctions) \cite{wiesenfeld91,nichols94,wiesenfeld96},
electrochemical oscillators \cite{wang00,kiss02c}, or
 continuous-flow stirred tank reactors containing yeast cells
\cite{dano99,dano01}. 
Globally coupled oscillators or maps also constitute a useful
approximation for a wider class of spatially extended systems, provided that
the range of the interaction is sufficiently large \cite{anteneodo03}.  
This may be the case for instance for swarms of flashing fireflies
\cite{winfree67,winfree80}, in networks of neurons \cite{sompolinsky91},
cultured heart cells \cite{soen99}, or mixed chemical reactions
\cite{neufeld03}.  

The interest in globally-coupled systems primarily arises from 
the a priori non-trivial relation between the dynamics of
one single element and the emergent behavior of the population, 
typically described by macroscopic variables. 
Given that these macroscopic quantities are often more easily 
---and sometimes exclusively--- accessible to 
experimental measurements, the non-trivial relation 
between the two description levels is a central problem 
not only in the modeling process, but also in the interpretation 
of experimental measurements.
A number of studies have dealt with the weak-coupling regimes 
of these noiseless systems. Particular attention has been paid to
the emergence of a macroscopic dynamics 
or ``non-trivial collective behavior''
\cite{kaneko89,kaneko90,chate91,maritan94,pikovsky94a,pikovsky94b,nakagawa98,chawanya98,manrubia00,shimada02,popovych01b,popovych02}.
Starting from the uncoupled regime, where the mean field is the average of an
infinite number of uncorrelated processes and therefore constant,
an increase in the coupling among
the population elements causes the macroscopic observables to depart 
from stationarity and to typically display complex collective
dynamics, such as clustering and multistability. 
In the strong-coupling limit, on the other hand, 
perfect synchronization of the individual elements is observed,
a regime of limited interest since the macroscopic evolution is then
a trivial copy of the microscopic one.

This limit is the point of departure for the present study of
the macroscopic dynamics of large populations
of globally and strongly coupled identical maps subjected to microscopic
disorder (additive noise or parameter mismatch). Indeed, starting from
perfect synchronization offers the hope that the effects of 
microscopic disorder can be 
disentangled from the subtleties of the collective regimes typical of
the weak-coupling regime.

Apart from the early work of Nichols and Wiesenfeld \cite{nichols94}
who studied how synchronous periodic dynamics is affected by noise
when the individual map is close to a bifurcation point, the only
prominent study of the effect of noise on the synchronous 
macroscopic dynamics of globally-coupled maps
has been conducted by Teramae and Kuramoto \cite{teramae01}. 
They showed that the unfolding of
a {\it chaotic} synchronous behavior by weak noise is generically
characterized by ``anomalous'' multiscaling properties of the cloud of points 
representing the population in the local phase space, in an approach typical
of that developed in a series of important papers by Kuramoto and Nakao on
non-locally 
coupled systems \cite{kuramoto96,kuramoto97a,kuramoto97b}.
Here we propose an alternative viewpoint on this indication
that microscopic additive noise interacts
non-trivially with the dynamics of the population.   

Starting from the perfect synchronization regime, we observe that the
collective motion remains 
deterministic and apparently low-dimensional at any noise level
(up to finite-size effects), at odds with the case where the added stochastic 
term is the same for every population element \cite{maritan94,pikovsky94b,toral01}. 
We present an analytical method for deriving the collective  
behavior from the single element dynamics and the statistical features of
the microscopic disorder \cite{demonte04a}. Expanding in series the
equation of motion of the ``mean-field'' (i.e. the average of our 
one-dimensional maps), we obtain an infinite hierarchy of equations for
macroscopic variables or order parameters. 
Our approach is independent of 
the specific form of the map and of the noise distribution. In
particular, the stochastic term needs not to be chosen neither weak nor
Gaussian. 
This method thus provides an answer to the problem of building
macroscopic equations of motion starting from the microscopic
structure of the population, as discussed e.g. by Cencini et
al. \cite{cencini99}, and 
justifies the use of specific systems as representatives of universality
classes of noise-induced phenomena. 
The proposed approach is complementary to the moment expansion method applied
in previous studies
\cite{desai78,pikovsky94a,topaj01,pikovsky02,lindner04}, where the macroscopic
equations are derived from the cumulant expansion of the Fokker-Planck or
Perron-Frobenius equation for the Langevin dynamics. Our approach
differs from the moment 
expansion techniques in two aspects: first, we do not start from the dynamics
of the probability distribution function, but directly from the individual
dynamics; second, we do not impose any self-consistency condition, but
closure of the equations relies on the dependence on the coupling strength.
On the other hand, the order parameter expansion is analogous in spirit to the
method recently proposed by Schimansky-Geier and coworkers
\cite{kawai03,lindner04}.

The paper is organized as follows.
In the next section, we introduce the phenomenology of {\it noise-induced
macroscopic bifurcations} of large populations of noisy chaotic maps. By means
of numerical simulations, we show how the addition of independent noise to
each map modifies the mean-field behavior, and we discuss the microscopic
features corresponding to different couplings and noise intensities.
This picture agrees with the phenomena of noise-induced macroscopic
bifurcations that have been studied in the context of continuous-time
dynamical systems \cite{rappel96,zaks03}.
Section\ \ref{sec:nope} is devoted to the derivation of the
finite-dimensional maps, or reduced systems, which approximate the mean-field
dynamics. The detailed analytical derivation through an expansion in order
parameters can be found in the appendices.
In Section\ \ref{sec:bifdiag} we compare the main features
of the collective dynamics of a population of chaotic maps 
to those exhibited by the reduced systems. 
In particular, we investigate the connection between our findings and the
anomalous scaling properties put forward by Teramae and Kuramoto.
Section\ \ref{sec:cfr} compares the effect of noise 
to that of parameter mismatch (``quenched disorder''), and discusses
the differences in deriving the reduced systems in the two cases 
\cite{demonte02,demonte03}.
Section\ \ref{sec:discuss} summarizes the main results of the paper,
discusses the perspectives of our method and points to a number of possible
applications. Part of our results have been presented in \cite{demonte04a}.

\end{section}

\begin{section}{Noise-induced collective regimes: phenomenology}
  \label{sec:phenomenology}  

Let us consider a population of globally-coupled maps in the formulation
introduced by Kaneko \cite{kaneko89}.
Each individual variable is ruled by the equation:
\begin{eqnarray} \label{eq:npop}
x_j\mapsto (1-K)\,f(x_j)+K\,\nsum{f(x)}+ \xi_j(t) \quad j=1,\dots N,
\end{eqnarray}
where $f: \mathbb R \rightarrow \mathbb R$ 
defines the single-element dynamics, which we require to be smooth
for our analytical approach,
and $\langle\dots\rangle$ denotes averaging over the entire population.
The stochastic term $\xi_j(t)$ is chosen at each time step, independently for
each population element, according to a distribution of zero average and with
the assigned moments $m_q$ (the second of which will be denoted $\sigma^2$).
The coupling strength $K$ ranges from zero, when the population elements are
decoupled, to one, when every element is mapped, in the absence of noise, into
the same average value. Note that these equations
can be recast into a form where the coupling constant can take any positive
value, a formulation that is commonly used in the case of globally coupled
continuous-time systems.  

Next, we use numerical simulations to
show how the evolution of the average state variable, 
the mean-field $X=\nsum{x}=\frac 1 N \sum_{j=1}^N x_j$,
is affected by a change in the noise intensity, as measured by the variance
$\sigma^2$ of the noise distribution. Together with the coupling strength $K$,
$\sigma^2$ will be taken as a control parameter for the analysis of the
macroscopic dynamics. 
As already mentioned, we focus on the strong coupling regime
$K\in[K_{\rm c},1]$ where the maps synchronize
perfectly in the absence of noise.

As one would expect by continuity, if only a small amount of noise is added,
the population configuration will be a small perturbation of the fully
synchronous regime: 
the time series of each individual element remains close to that of the 
mean-field, even in chaotic regimes.
This is illustrated in Fig.~\ref{fig:nts}(a) for a population 
of $2^{22}$ logistic maps of the form $f(x)=1-a\,x^2$
with the nonlinearity parameter $a=1.57$ 
chosen inside the chaotic region, subjected to uniformly distributed
noise of zero mean.
(All examples presented in this
paper will refer to a population of logistic maps, 
while the application of our approach to
populations with different microscopic features will be addressed elsewhere.)
When, instead, the noise is strong, the coherence of the motion within the
population is lost. The evolution of any element is now blurred by the
stochastic term, thus making it difficult to recognize 
any underlying deterministic behavior from the detection 
of just one individual time series.     
Up to finite-size effects, however, the mean field evolves deterministically. 
The regimes of macroscopic dynamics induced by noise are neither synchronous,
since the dynamics of the individual elements are not locked to each other, 
nor forcedly coherent, in the sense that the distances between units
and the mean-field are not always small.   

\begin{figure}[h]
\center
\epsfig{file=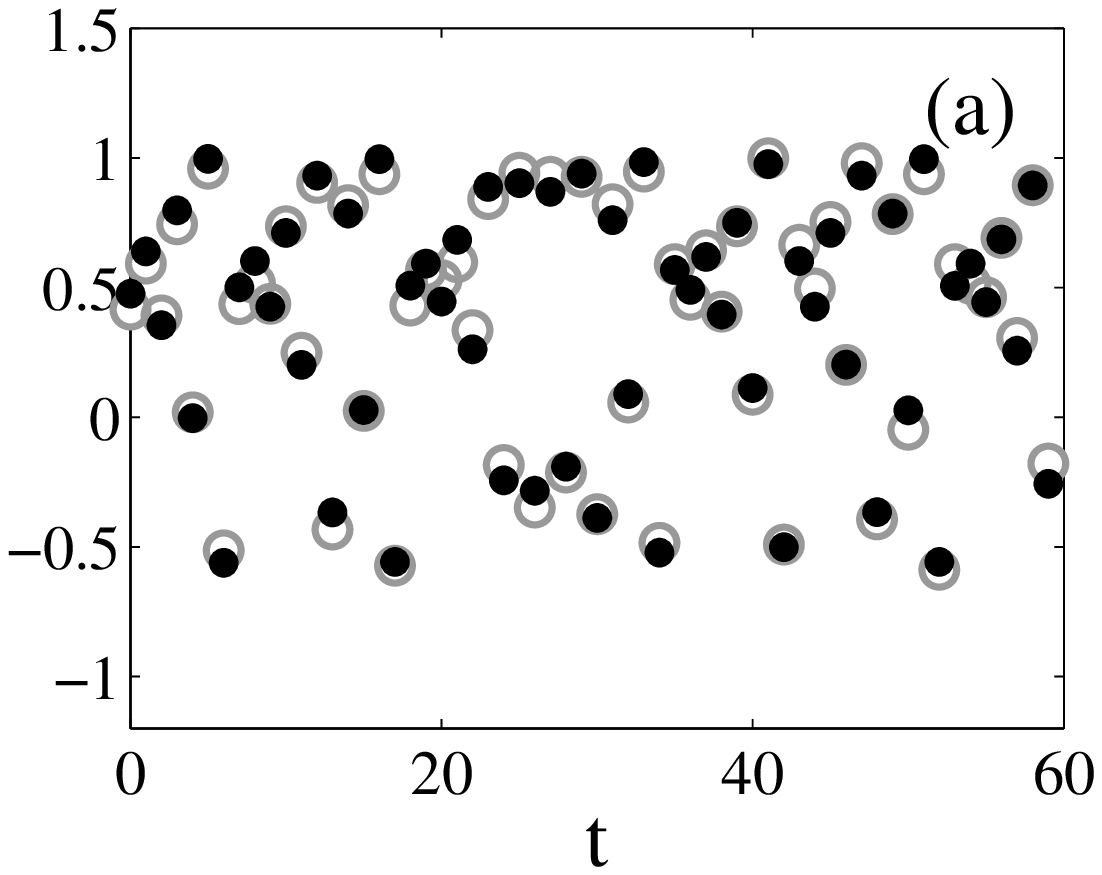, width=.45\textwidth}
\epsfig{file=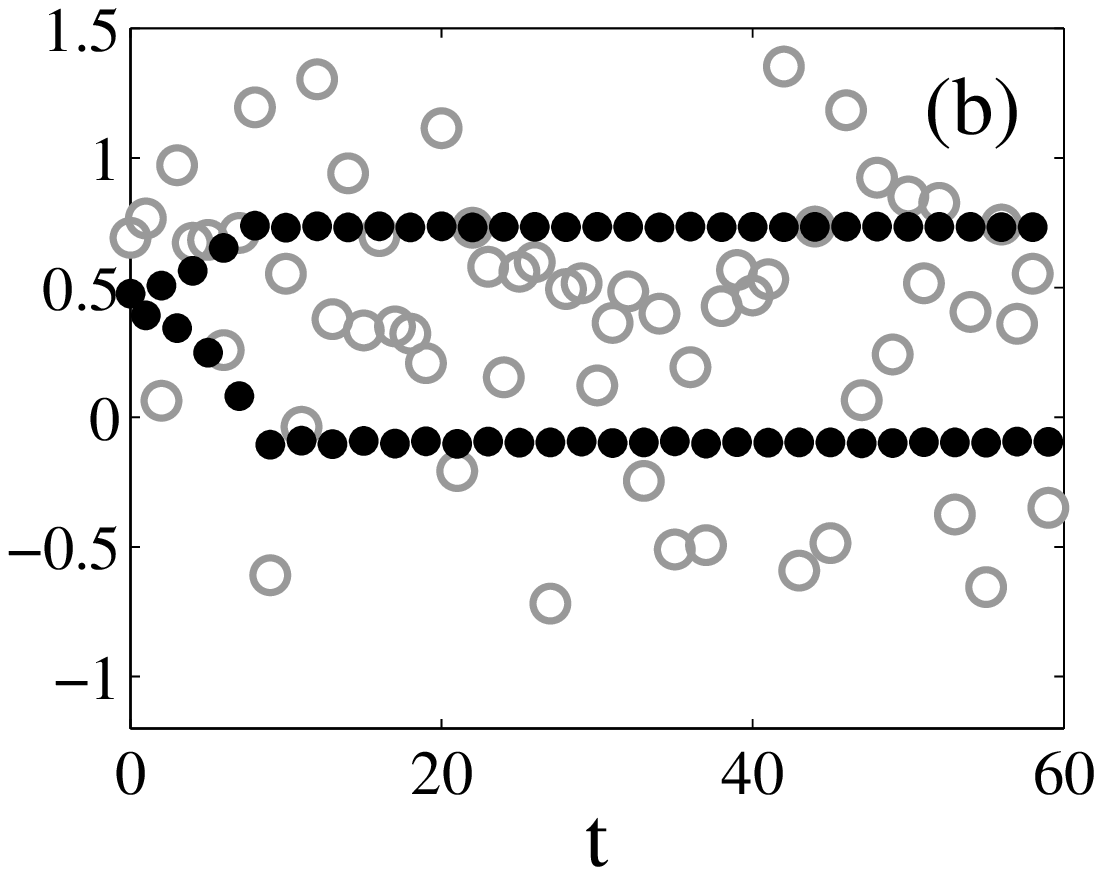, width=.45\textwidth}
\caption{Time series of the mean-field X (black dots) and of one
  element of the population (open circles) for $K=1$. Population of
  $N=2^{22}$ logistic maps in the chaotic region ($a=1.57$) with
  uniformly distributed noise, random initial conditions. (a) In the case
  of weak noise 
  ($\sigma^2=0.05$), the dynamics is a small perturbation of the
  synchronous, chaotic, noiseless one; (b) In the case of stronger noise
  intensity ($\sigma^2=0.4$), the mean field attains a regular, periodic
  behavior   while the time series of the single population element is
  scattered by noise.\label{fig:nts}}  
\end{figure}

It is remarkable that, when the noise is sufficiently intense, not only does
the mean-field evolution appear to remain 
low-dimensional, but it can be qualitatively different from the
single-element uncoupled dynamics.  
Figure\ \ref{fig:nts}(b) shows for instance that, in the presence of
microscopic noise, the mean-field can display regular cyclic behavior in
spite of the fact that every element of the population is itself chaotic
and noisy.
This phenomenon is related to the fact that noise leads the trajectory of any
individual element to stay away from the chaotic attractor, so that its
dynamics is 
mainly influenced by the structure (attractive and repulsive manifolds, basins
of attraction) of the single-element phase space in the proximity of the
asymptotic solution. The coupling among the elements then let these
different processes interact so that the probability for one
individual system to be mapped into a particular region of its phase space
is larger if many other individuals are mapped into similar regions. By
averaging over the 
whole population, the emergent collective dynamics will be mainly determined
by how the phase space is structured and, thus, by the global features of the
single-element dynamics.

Between the two cases shown in Fig.\ \ref{fig:nts}, a series of other
macroscopic regimes is observed, 
ranging from (one or two-band) chaos to cycles of different periods. This
can be summarized in the bifurcation 
diagram of the mean-field shown in Fig.~\ref{fig:newfig}(a).
(For even higher noise intensities, divergences occur if the single-element
map is defined in an interval, as is the case of the logistic map considered
here. For unbounded noise distributions, one should in principle consider
maps of an unbounded variable.) 

\begin{figure}[h]
\center
\epsfig{file=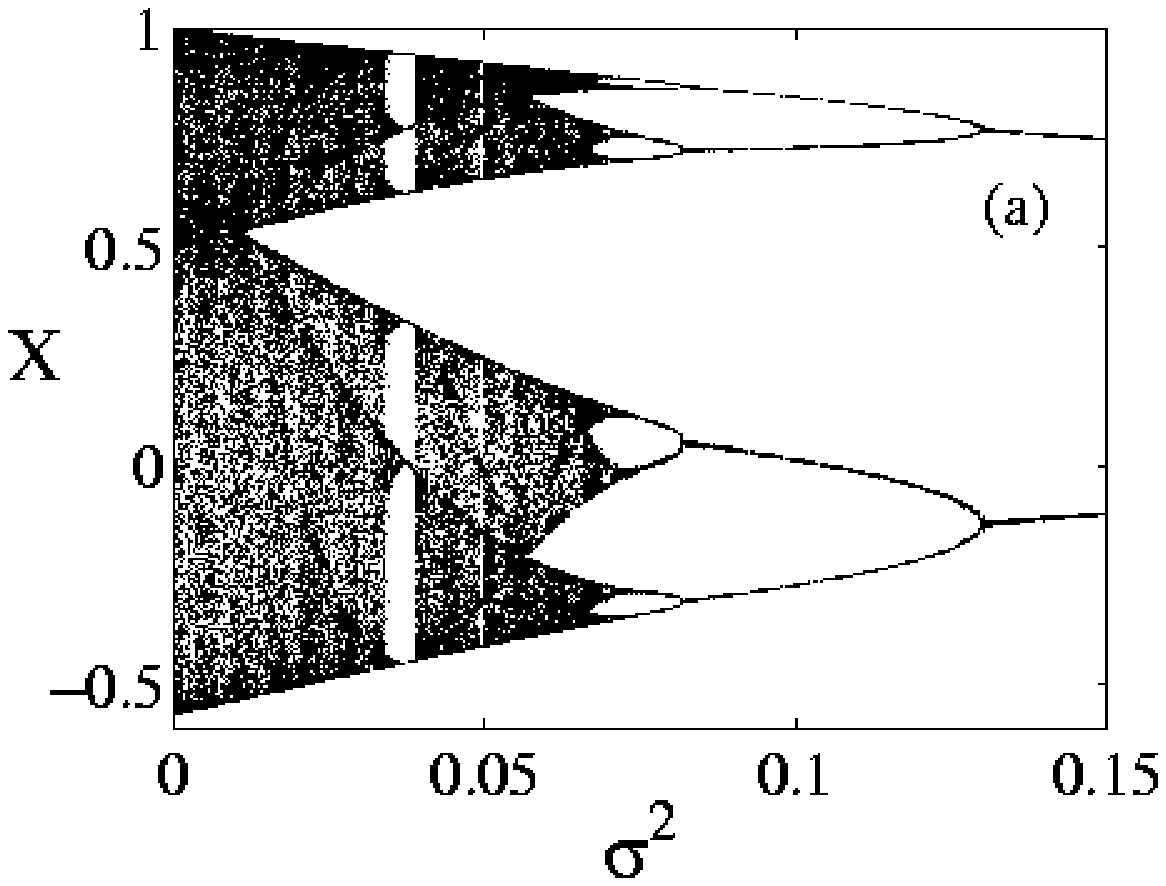, width=.48\textwidth}\hfill
\epsfig{file=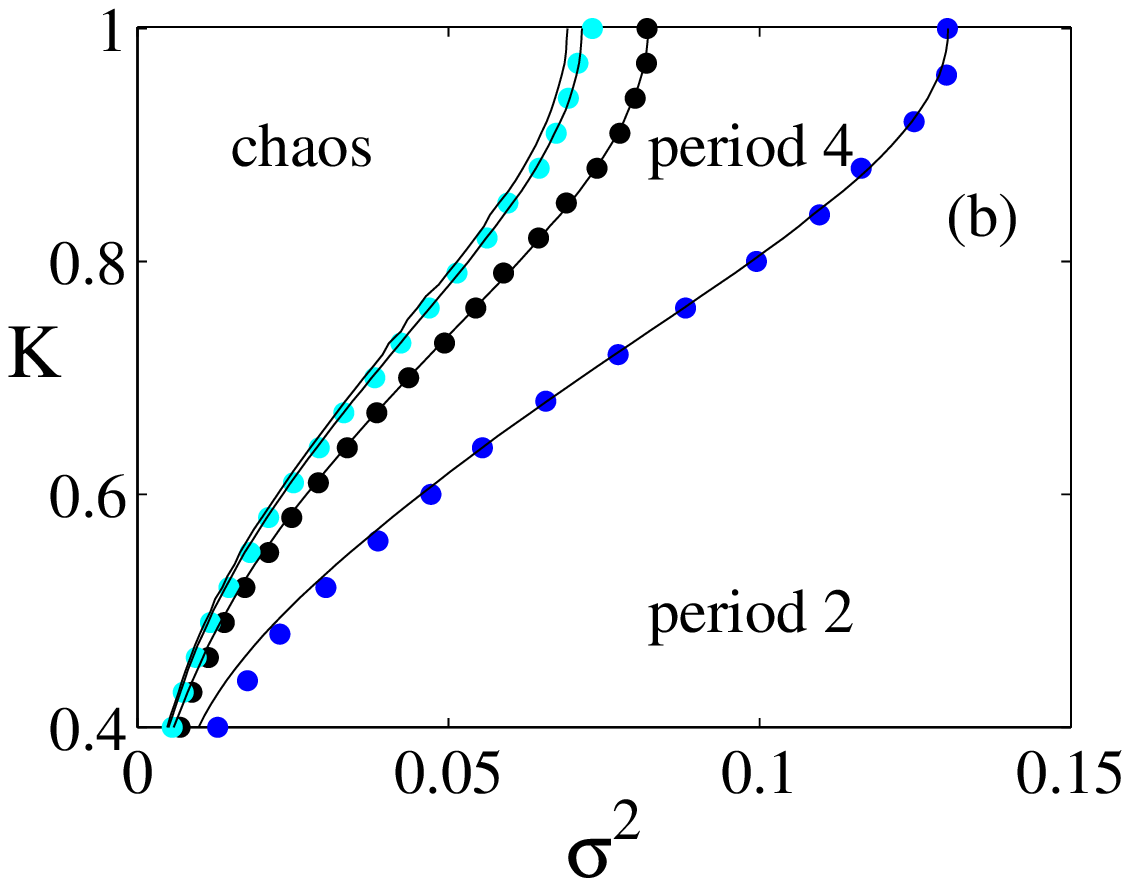,width=.47\textwidth}
\caption{Globally coupled logistic maps ($a=1.57$) 
with additive uniform noise.
(a) Bifurcation diagram of the mean-field
as $\sigma^2$ is varied at maximal coupling $K=1$.
(b) Phase diagram in the $(K,\sigma^2)$ plane, indicating the first
bifurcation lines of the period-doubling cascade (dots: full system, lines:
the reduced system to the second degree Eq.\~(\ref{eq:log2})).}
\label{fig:newfig} 
\end{figure}

In spite of its resemblance with the bifurcation diagram
of a single element, it is important to remember that the 
macroscopic transitions observed in Fig.~\ref{fig:newfig}(a) 
are directly caused by the presence of noise, 
while the population parameters are left unchanged.
Nonetheless, this macroscopic bifurcation diagram suggests that a
low-dimensional map might exist, whose bifurcations reproduce
the qualitative changes of the mean-field dynamics. Section~\ref{sec:nope}
IS dedicated to the derivation of such an effective dynamical system. 
Note further that the observed macroscopic bifurcations are similar to
those observed for globally coupled maps with stochastic updating
 \cite{morita02}, where the uncertainty on the mean field that every element
experiences is given by the nature of the updating rather than by an explicit
noise term, and for coupled map lattices
\cite{lemaitre96,lemaitre97,lemaitre99}, 
where the fluctuation of the local field are due to the existence of a finite
correlation length.
%what is the difference? if any

The microscopic and statistical features of these
noise-induced regimes are revealed by looking at three kinds of 
probability density functions (pdfs): (i) the population (or snapshot)
pdf, relative to the values assumed by the individual maps at a specific time,
(ii) the individual pdf and (iii) the mean-field pdf, computed over a large
time interval for one map of the population and for the mean field,
respectively.
In the regime of perfect synchronization, the snapshot pdf is a delta peak
centered on the value of the mean field, while the two time-averaged
pdfs coincide and give the probability measure of the single-element chaotic
attractor. 
When noise is increased, the pdfs are in general modified in a nontrivial
manner by the 
interplay of noise, that tends to blur the individual dynamics and broaden
both the population and the individual pdfs, and coupling,
that maintains a degree of coherence within the population.
As a first example, we consider the case where the coupling is maximal
and the mean-field displays a period-two cycle, as in Fig.~\ref{fig:nts}(b). 
It is easily seen from Eq.~(\ref{eq:npop}) that in this limit case 
the population is distributed exactly like the stochastic term. 
Accordingly, Fig.~\ref{fig:npdfs}(b) shows that the population of logistic
maps previously considered is at any time step uniformly
distributed around the mean field (not shown).
This is also reflected in the individual pdf if this is computed at even
and odd times separately (Fig. \ref{fig:npdfs}(a)). 
The fact that
the individual pdf is slightly broader than the instantaneous pdf and that the
mean field pdf is not exactly concentrated on the values 
taken during the deterministic motion is a consequence of finite size
effects. These effects introduce fluctuations that vanish for $N\to
\infty$.

\begin{figure}[h]
\center
\epsfig{file=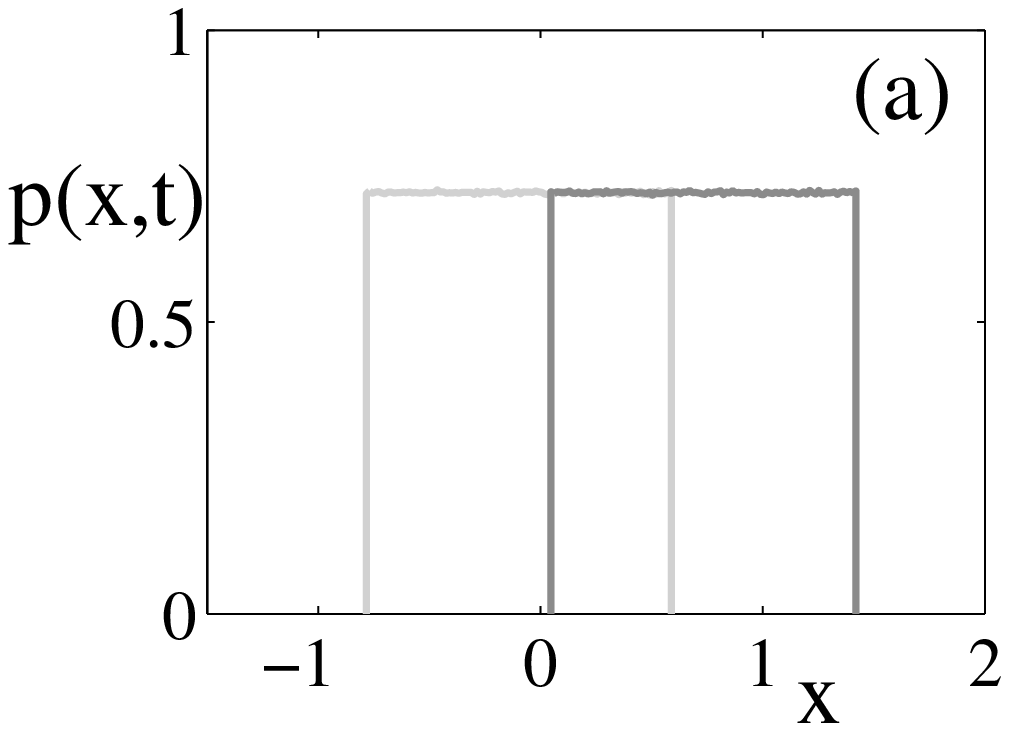, width=.3\textwidth}
\epsfig{file=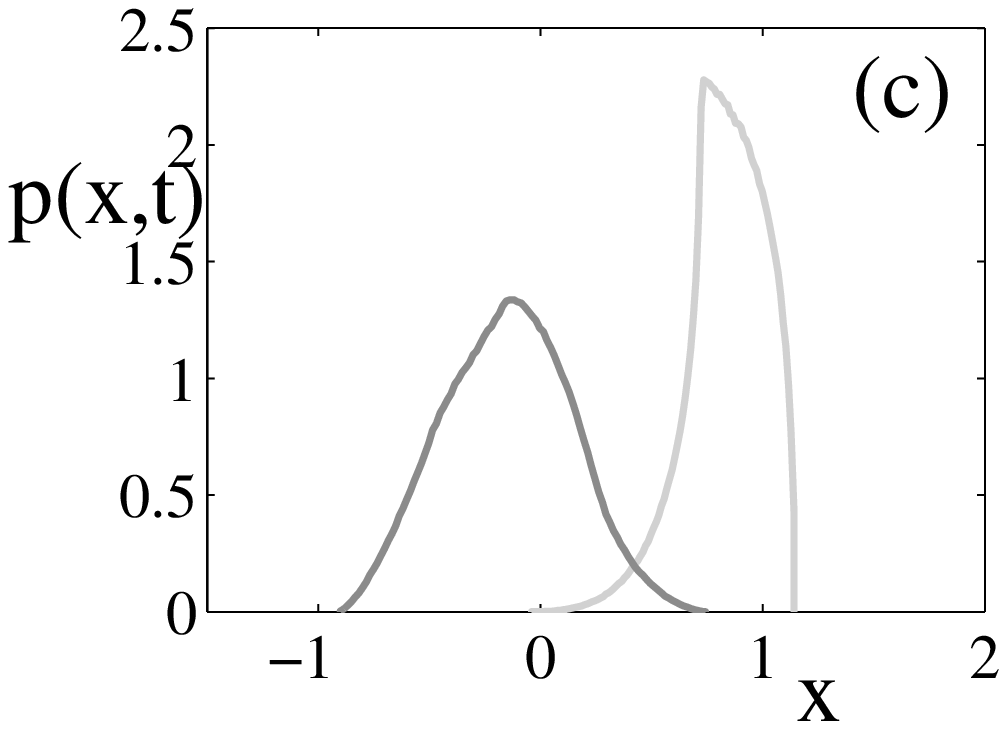, width=.3\textwidth}
\epsfig{file=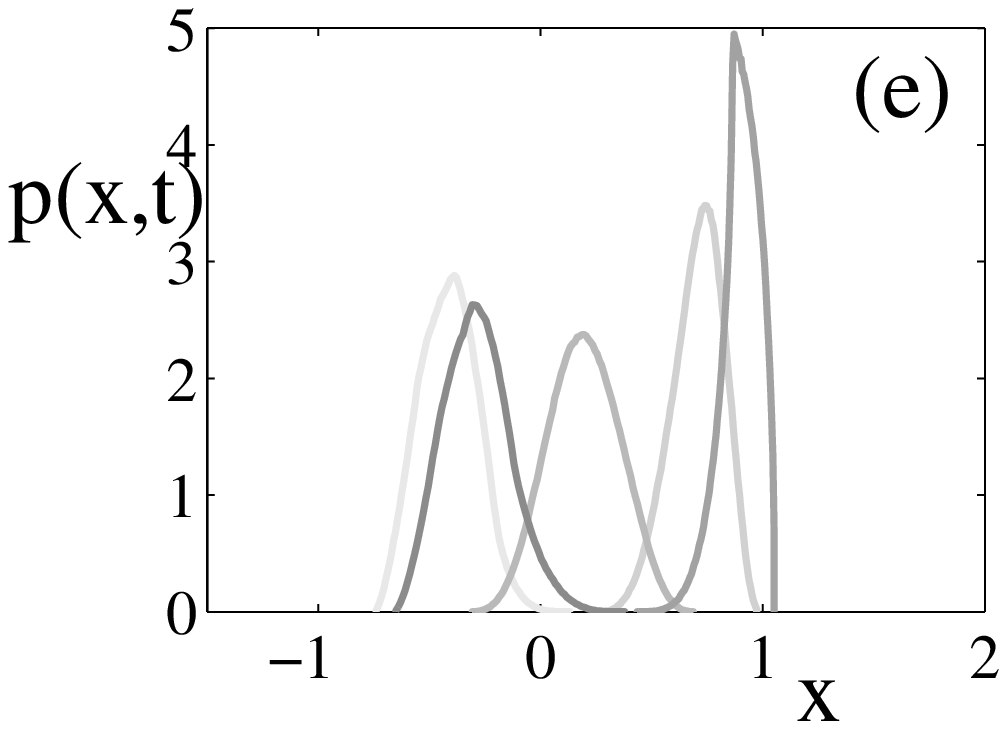, width=.3\textwidth}

\epsfig{file=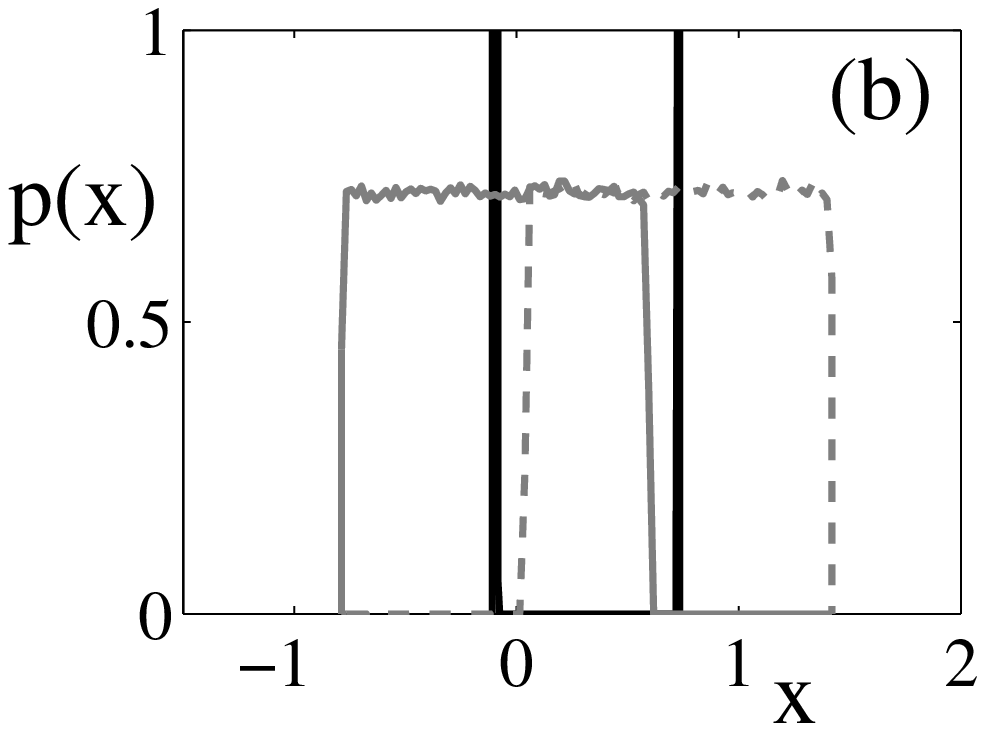, width=.3\textwidth}
\epsfig{file=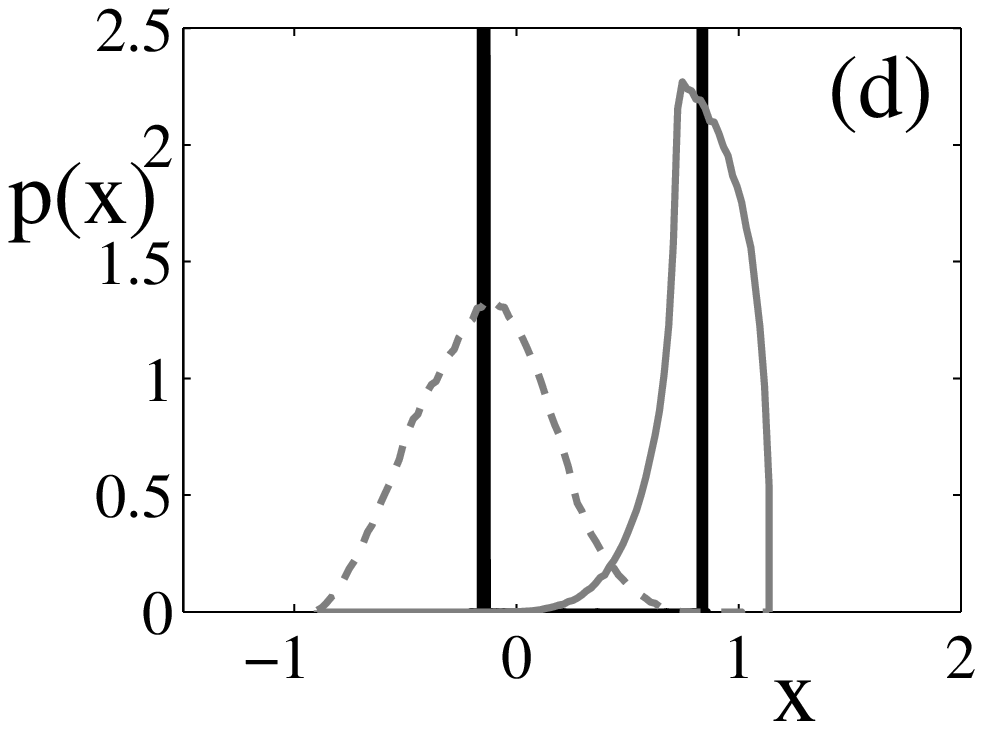, width=.3\textwidth}
\epsfig{file=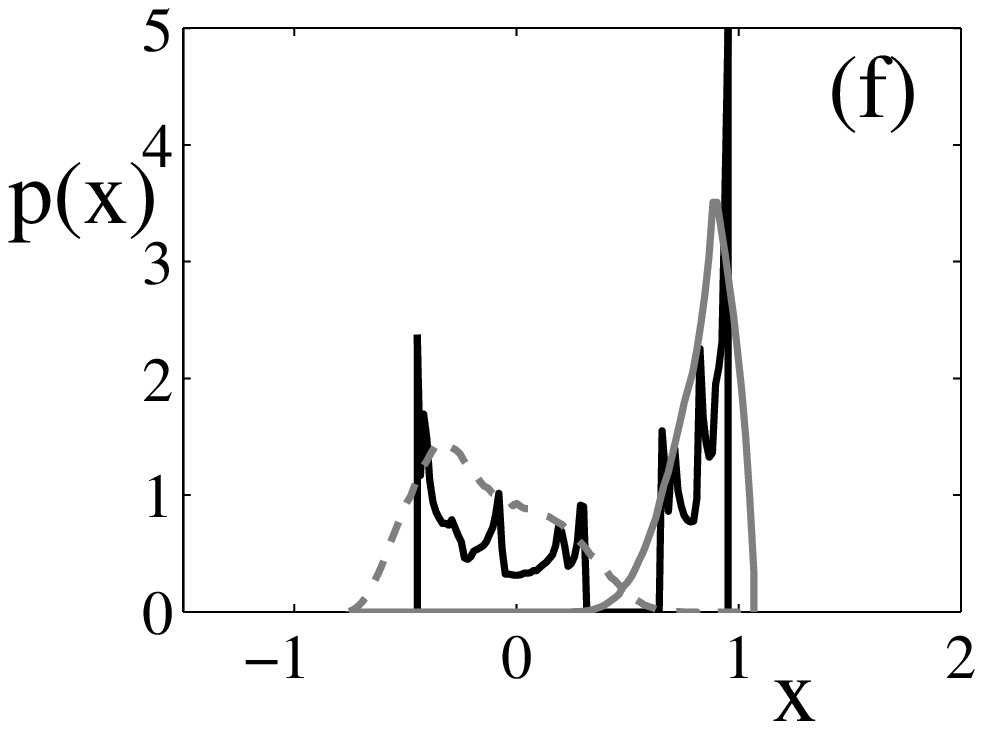, width=.3\textwidth}

\caption{Snapshot pdfs for two (a,c) and five (e) successive time steps and
  mean field and individual pdfs (b,d,f) for the population of Figure
  \ref{fig:nts}:
  (a-b) Maximal coupling $K=1$ and noise variance $\sigma^2=0.016$.
  (c-d) intermediate coupling $K=0.4$ and noise intensity
  $\sigma^2=0.015$. The mean 
  field displays a period-two cycle, as in (a-b):
  The interplay between coupling and nonlinearity of the single-element
  dynamics leads to a nontrivial reshaping of the invidual pdf, that no longer
  reflects the shape of the noise distribution.
  (e-f) intermediate coupling $K=0.4$ and noise intensity
  $\sigma^2=0.003$. 
  \label{fig:npdfs}}
\end{figure}

Let us now consider the case in which the coupling is not maximal, but still
strong enough to ensure perfect synchronization in the absence of noise. For
the population considered here, $K=0.4$ is right above the region where
two-cluster dynamics occurs.
Again, an increase in the noise intensity leads to a seemingly low-dimensional
bifurcation diagram. The amount of noise necessary to let the system reach a
periodic regime becomes smaller as the coupling strength is decreased.
If $K$ is sufficiently weak, a folded structure becomes visible in the
first return map of the mean-field, indicating that
the macroscopic dynamics is actually embedded in a space of dimension
greater than one.
The microscopic signature of the system is also significantly changed for
intermediate coupling strengths.
If we consider $K=0.4$ and $\sigma^2=0.015$, where the mean field displays a
period-2 cycle, like in the case of maximal coupling discussed above, we
notice immediately that the correspondence between the instantaneous pdf and
the noise distribution is lost (Fig.\ \ref{fig:npdfs} (b) and (c)).
The individual pdfs for odd/even times still reflect the instantaneous pdf
and the mean-field pdf is peaked around the values assumed during the cycle.  
If the noise intensity is also reduced, so that both the coupling and
the noise intensity are sufficiently weak, the possibility of inferring
the individual pdf from the instantaneous distribution is lost. For
$\sigma^2=0.003$, indeed, the mean field displays two-band chaos.
Figures\ \ref{fig:npdfs} (e) and (f) shows that the population distribution
changes 
in time depending on the actual position of the mean field on the macroscopic
attractor. 
The individual pdfs for odd/even times still allow us to recognize a
periodicity in the motion, due to the presence of two bands, while the mean
field pdf corresponds to the probability measure of the macroscopic attractor.

The phase diagram of our system in the $(K,\sigma^2)$ parameter plane
is summarized in Fig.~\ref{fig:newfig}(b).
For sufficiently large noise
intensity and coupling strength, the macroscopic 
dynamics is cyclic and of period two. By decreasing the noise intensity or
increasing the coupling (for sufficiently low noise) the macroscopic dynamics
undergoes a period-doubling bifurcation cascade, leading to regimes of
collective chaos.
For coupling strength weaker than those displayed in Fig.~\ref{fig:newfig}(b),
the system shows clustering and
multistability. Moving towards the limit of the
synchronous (noiseless) regime, the population pdf becomes bimodal or
multimodal and its moments take values significantly larger than
those of the noise distribution.
This transition to clustered solutions, where the population is divided into
subgroups placed at finite distance, is related to the 
anomalous scaling properties reported by Teramae and Kuramoto,
see Sec.~\ref{sec:bifdiag}. 
We note finally that these findings are in agreement with those of
Shibata, Chawanya and Kaneko \cite{shibata99a} who observed that very weak
noise is sufficient to drastically
reduce the dimensionality of the macroscopic dynamics at weak coupling.
\end{section}

\begin{section}{Order parameter expansion}\label{sec:nope}

In this section we present our order parameter expansion and its
truncation to finite-dimensional systems.
The method is formally analogous to that formulated 
in our recent publications \cite{demonte02,demonte03} for populations with 
parameter mismatch, but relies on different closure assumptions based on the
different microscopic properties of the noisy system. 
These differences and their implications will be discussed
in Sect.~\ref{sec:cfr}.
We only present here the main ideas behind our derivation and refer
the reader to Appendix A for details.

The iterate of the mean-field can be formally computed by averaging the
individual dynamics Eq.\ (\ref{eq:npop}) and by performing the change of
variables: $x_j=X+\epsilon_j$. In this way, the dynamics of the oscillators
around the mean-field is decoupled from the deterministic behaviour of their
average. 
By expanding in series the single-element map $f$ 
{\it around the mean-field}, one obtains:
\begin{eqnarray}\label{eq:X}
X\mapsto \nsum{f(x)}=
f(X)+\sum_{q=1}^\infty\:\frac{1}{q!}\:\ma{D}^qf(X)\:\nsum{\epsilon^q}+
\nsum{\xi}
\end{eqnarray}
where $\ma{D}^pf$ is the $p$-th derivative of the single-element dynamics,
and the last term, reflecting finite-size fluctuations, vanishes
in the infinite-size limit for zero-mean noise distributions. 
Note that $X$ is coupled to other macroscopic variables, namely to the
moments of the population instantaneous pdf, defined by:
\begin{eqnarray}\label{eq:ordpar}
\Omega_q:=\nsum{\epsilon^q}\hspace{60mm}q\in\mathbb N.
\end{eqnarray}
From now on we will refer to these macroscopic observables as \emph{order
  parameters}. 
This is not only to avoid confusion with the noise distribution 
moments, but also to stress that these are the degrees of freedom relevant for
a macroscopic description. Moreover, as will be shown in 
Sec.~\ref{sec:cfr}, in the more general case in which the 
elements are not identical, the same expansion leads to order parameters that
do not coincide with the moments of the population pdf.      

The iterates of the order parameters can be computed by following the same
scheme of variable change and averaging the iterates of the displacements
$\epsilon_j$, obtained after a Taylor expansion of the individual maps. 
This yields:
\begin{eqnarray*}
\Omega_q\mapsto &&
\sum_{i=0}^q \binom{q}{i} (1-K)^i
\left\langle \left(\xi-\nsum{\xi}\right)^{q-i}
\left[\sum_{p=1}^\infty\frac{1}{p!}\ma{D}^pf(X)
\left(\epsilon^p-\Omega_p\right)\right]^i\right\rangle\;. \nonumber
\end{eqnarray*}
Observing that, as a consequence of the fact that the displacements 
$\epsilon_j$ and the noise are uncorrelated variables, 
$\nsum{h(X,\epsilon)\:\xi^q}=\nsum{h(X,\epsilon)}\:\nsum{\left(\xi-\nsum{\xi}\right)^q}$, and taking the limit $N\to \infty$, the equations for the order
parameters become:
\begin{eqnarray}\label{eq:omega}
\Omega_q\mapsto && 
m_q+ \sum_{i=1}^q \binom{q}{i} (1-K)^i\:m_{q-i}\nonumber
\left\langle\left[\sum_{p=1}^\infty\frac{1}{p!}\ma{D}^pf(X)
\left(\epsilon^p-\Omega_p\right)\right]^i\right\rangle,
\end{eqnarray}
where $m_q=\nsum{\left(\xi-\nsum{\xi}\right)^q}$ is the $q$-th moment of the
noise distribution. 

Such equations 
compose an infinite-dimensional dynamical system describing the evolution
of all order parameters, formally equivalent, in the infinite-size limit,
to the original system.
The dependence of these equations on powers of $(1-K)$ allows us
to truncate them in the region of strong coupling. We call
such a truncation to the $n$-th power in $(1-K)$ 
the {\it reduced system of $n$-th
degree}.
In Appendix\ \ref{app:ope} we demonstrate that for any polynomial map of
degree $P$ the reduced system at $n$-th degree is a map 
with $n$ independent variables (the mean field $X$ and the order parameters
form the second to the $n$-th) and $(n-1)P$ slaved variables (the order
parameters from number $n+1$ to number $n\:P$). The remaining order
parameters are constantly equal to the moments of the noise distribution.  
The reduced system to the $n$-th degree can thus be cast in the form:
\begin{eqnarray}\label{eq:redpol}
\begin{cases}
X\mapsto&f(X)+\sum_{q=2}^n\:A_q(X)\Omega_q 
\\\rule[0mm]{0mm}{7mm}
\Omega_q\mapsto &
m_q+ \sum_{i=1}^q \binom{q}{i} (1-K)^i\,m_{q-i}
\,\Gamma_i (X,\Omega_2, \dots \Omega_n) \hspace{7mm} q=2,\dots n
\end{cases}
\end{eqnarray}
where the functions $A_q$ and $\Gamma_i$ of the
mean-field and of the order parameters relevant to the chosen approximation
level contain the first $2P$ moments of the noise distribution as
population-level parameters. 
As  illustrated in Appendix~\ref{app:ope}, such expressions can be
systematically derived in two steps: first, one writes the order parameters 
$\Omega_{n+1} \dots \Omega_{nP}$ as linear combinations of $\Omega_2 \dots
\Omega_n$, then these are plugged into Eq.\ (\ref{eq:X}) and into the first
$n$ terms of Eq.\ (\ref{eq:omega}). 

As a first example, consider the reduced system of zeroth degree:
\begin{eqnarray}\label{eq:X0}
X\mapsto 
f(X)+\sum_{q=1}^\infty\:\frac{1}{q!}\:\ma{D}^qf(X)\:m_q.
\end{eqnarray} 
This scalar map accounts exactly for the macroscopic dynamics when the
coupling is  maximal (see, e.g. Fig.~\ref{fig:newfig})
and provides a first approximation for the dynamics at
very strong coupling. 

Equation\ (\ref{eq:X0}) naturally contains the interplay
between nonlinearities of the uncoupled map, represented by the derivatives of
the single-element dynamics, and the features of the noise distribution, given
by its moments.
In particular, it allows us to infer that, if the single-element dynamics is
polynomial, only a finite number of the noise distribution moments $m_q$ will
influence the macroscopic dynamics. 
This induces a relation of equivalence onto
the space of the distributions for the stochastic term. These distributions can
thus be divided into classes on the bases of their effect on the mean-field.

Being independent of $K$, the reduced system to zeroth degree obviously
cannot describe the change in the bifurcation values when the coupling is
lowered, and higher degree truncations need to be considered.
With zero-average symmetric noise distributions (such as envisaged here),
only even-degree order parameters come into play.

The reduced system to the second degree is the first truncation that displays
an explicit dependence on the coupling constant $K$. 
Such a truncation differs from the Gaussian approximation of the population
pdf, commonly used for closing cumulant expansions \cite{zaks03}, by which the
population pdf is approximated as being completely characterized by its
first two moments. Our closure assumptions are instead based on the strength
of the coupling and take into account also for cases in which neither the noise
distribution nor the snapshot pdfs are Gaussian.
However, in the truncations at lowest degree, the equations obtained with the
two different assumptions may coincide. 

In order to compute the two-dimensional reduced system, we exploit the
recurrence relation for the order parameters:
\begin{eqnarray*}
\Omega_{q+2}=m_{q+2}-\frac{(q+2)(q+1)}{q(q-1)}\frac{m_q}{m_{q-2}}(m_q-\Omega_q)
\end{eqnarray*} 
that can be derived from Eq.\ (\ref{eq:appgamma}).
This allows us to express every order parameter as a function of $\Omega_2$
through the computation of a telescopic sum, that yields:
\begin{eqnarray}\label{eq:oq_o2}
\Omega_q=m_q-\frac{q(q-1)}{2}\:m_{q-2}\:\left(\sigma^2-\Omega_2\right).
\end{eqnarray}
One of the consequences of the last equation is that all the odd order
parameters are given by the corresponding moments of the
noise distribution. 
\\
Equation\ (\ref{eq:omega}) can be rewritten, expliciting the squared term, as: 
\begin{eqnarray*}
\Omega_q\mapsto&& m_q+ \frac{q(q-1)}{2}\:(1-K)^2\:m_{q-2}
\sum_{p,r=1}^\infty \frac {1}{p!\:r!} 
\:\left[\ma{D}^p\,f(X)\right]\:\left[\ma{D}^r\,f(X)\right]\:\Omega_{p+r}\\
&&-\frac{q(q-1)}{2}\:(1-K)^2\:m_{q-2}\left[\sum_{p=1}^\infty \frac {1}{p!} 
\:\ma{D}^p\,f(X)\:\Omega_p\right]^2.
\end{eqnarray*}
Substituting in this last equation the recurrence Eq.\
(\ref{eq:oq_o2}), one gets the two-dimensional reduced system:
\begin{eqnarray}\label{eq:ope_2}
\begin{cases}
X\mapsto& 
f(X)+\alpha_0(X)+\alpha_1(X)\:\Omega_2\\\rule[0mm]{0mm}{6mm}
\Omega_2\mapsto& \sigma^2+\:(1-K)^2 \left[ \gamma_0(X)
+\gamma_1(X)\:\Omega_2
+\gamma_2(X)\:\Omega_2^2 \: \right].
\end{cases}
\end{eqnarray}
The nonlinear dependence from the mean field is contained into the
coefficients: 
\begin{eqnarray*}
&&\alpha_0(X)=\sum_{q=2}^\infty\:\frac{1}{q!}\:\ma{D}^qf(X)
\left[\:m_q-\binom q 2\:m_{q-2}\:\sigma^2\right]\\
&&\alpha_1(X)=\frac 1 2 \sum_{q=2}^\infty\:\frac{1}{(q-2)!}\:
\ma{D}^qf(X)\:m_{q-2}\\
&&\gamma_0(X)=-\left[\alpha_0(x)\right]^2+\beta_0(X)-\beta_1(X)\:\sigma^2\\
\rule[-2.5mm]{0mm}{8mm}
&&\gamma_1(X)= \beta_1(X)-2\:\alpha_0(X)\,\alpha_1(X)\\
&&\gamma_2(X)=-\left[\alpha_1(X)\right]^2,
\end{eqnarray*}
with:
\begin{eqnarray*}
&&\beta_0(X)=\sum_{q,p=1}^\infty \frac {m_{q+p}}{q!\:p!}\:
\left[\ma{D}^q\,f(X)\right]\:\left[\ma{D}^p\,f(X)\right]\\
&&\beta_1(X)=\sum_{q,p=1}^\infty \frac
{m_{q+p-2}}{q!\:p!}\binom {q+p} 2\left[\ma{D}^q\,f(X)\right]\:\left[\ma{D}^q\,f(X)\right] 
.
\end{eqnarray*}
These terms originate according to the nonlinearities of the single-element
dynamics and to the moments of the noise distribution.
If $f$ is a polynomial of degree $P$, the highest moment of the noise
distribution that occurs as a population-level parameter is $m_{2P}$, while the
moments of higher orders are unimportant to this approximation level.

The reduced systems obtained by successive approximations of the order
parameter expansion provide the link between the microscopic properties of the
population and the mean field dynamics. In the following section, we only
consider truncations up to the fourth 
degree for quadratic maps, but an algorithm can easily be implemented for
computing the reduced system at $n$-th degree for an 
arbitrary polynomial map.

The formalism introduced here allows us to conclude that the higher the
nonlinearity of the single-element dynamics is, the stronger the noise
intensity must be to observe its effect on the macroscopic dynamics. 
A second important conclusion that can be drawn from the order parameter
expansion is that the microscopic map determines which of the noise
distribution moments the mean field behaviour is sensitive to. These two
facts suggest that even if only averaged observables are accessible,
information about the features of the single-element dynamics can be inferred
from  purely macroscopic measurements.
\\

\end{section}

\begin{section}{Macroscopic bifurcation diagram and anomalous
    fluctuations.}\label{sec:bifdiag} 

In this section we show that the reduced systems are able to capture
the main properties of the collective dynamics. In fact, as will be 
reported elsewhere, the quality of the agreement goes well beyond these
main properties, and the reduced systems account, as their degree is
increased, for finer and finer features of the collective dynamics.

Figure~\ref{fig:newfig}(b) displays the bifurcation lines of the reduced
system of second degree for the population of logistic maps considered
in the previous section.  In this case
Eq.~(\ref{eq:ope_2}) is the two-dimensional map:
\begin{align}\label{eq:log2}\begin{cases}
X\mapsto & 1-a\,X^2-a\:\Omega_2\\
\Omega_2\mapsto & \sigma^2+(1-K)^2\:a^2
\left[m_4-6\,\sigma^4+(4X^2-\Omega_2+6\:\sigma^2)\:\Omega_2\right].
\end{cases}
\end{align}
The dependence of the reduced system on the variance and kurtosis of the
noise distribution, as well as on the coupling strength, is seen to reflect
the qualitative changes in collective behaviour that are observed for the
population. 

For low coupling intensities, that is approaching the region where the
synchronous regime becomes unstable for the noiseless maps, the
agreement between the reduced and the full system  deteriorates
up to the point where the solutions of Eq.~(\ref{eq:log2}) diverge. 
One then needs to consider higher-degree truncations, which indeed,
account better for the actual dynamics. For the large-coupling region
shown in Fig.~\ref{fig:newfig}(b), however, the improvement is 
hardly visible and the second-degree reduced system provides a
satisfactory description of the collective dynamics.

Strikingly, the quality of the second-degree reduced system is such that
this simple two-dimensional map also accounts for the onset of the anomalous
scaling properties uncovered by Teramae and Kuramoto \cite{teramae01}. 
In Fig.~\ref{fig:scalk}(a) we first show that anomalous scaling is 
observed for globally coupled noisy logistic maps, in agreement with 
the claims of universality made by these authors. Note,
though, that anomalous scaling is only observed over a {\it finite}
range of noise strengths:
the anomalous behaviour sets in for sufficiently large, although weak,
noise, and disappears for noise fluctuations of order one, i.e.
of the order of the size of the attractor.
We interpret the {\it normal} scaling observed at very weak noise 
as being due to the fact that the individual pdf is 
maintained below the critical scale that separates microscopic 
from macroscopic chaos, as defined by Cencini et al. in \cite{cencini99}. 
Below this scale, noise does not
alter the structure of the macroscopic phase space, so that normal scaling is
observed. When the cloud of points becomes large enough to cause, for
instance, the connection of two neighboring folds of the noiseless system
invariant manifolds, the mean-field attractor undergoes a macroscopic
bifurcation. We trace the origin of anomalous scaling back to such
macroscopic (albeit possibly infinitesimally small) noise-induced changes
which are typically accompanied by intermittent behavior.
As already mentioned, the reduced system to second degree is 
sufficient to explain the appearance of anomalous fluctuations 
in the second moment of the individual pdf, i.e. the order parameter
$\Omega_2$. 
In Figs.~\ref{fig:scalk}(b,c), we compare the full and the reduced
system by plotting the ratio between the second moment of the
instantaneous distribution for the population and  $\sigma^2$,
the variance of the noise. The agreement is quantitative, and improves
in the region of low coupling when
the reduced system to fourth degree is considered (not shown).
It is not clear whether  our finite-dimensional reductions of the
full infinite-dimensional system are able to exhibit anomalous {\it scaling};
indeed the region where scaling is observed over many decades is limited
to the proximity of the transition to clustering/loss of full synchrony,
where our low-degree approximations fail.

\begin{figure}[h]
\center
\epsfig{file=anomalous.eps,width=.85\textwidth}

\epsfig{file=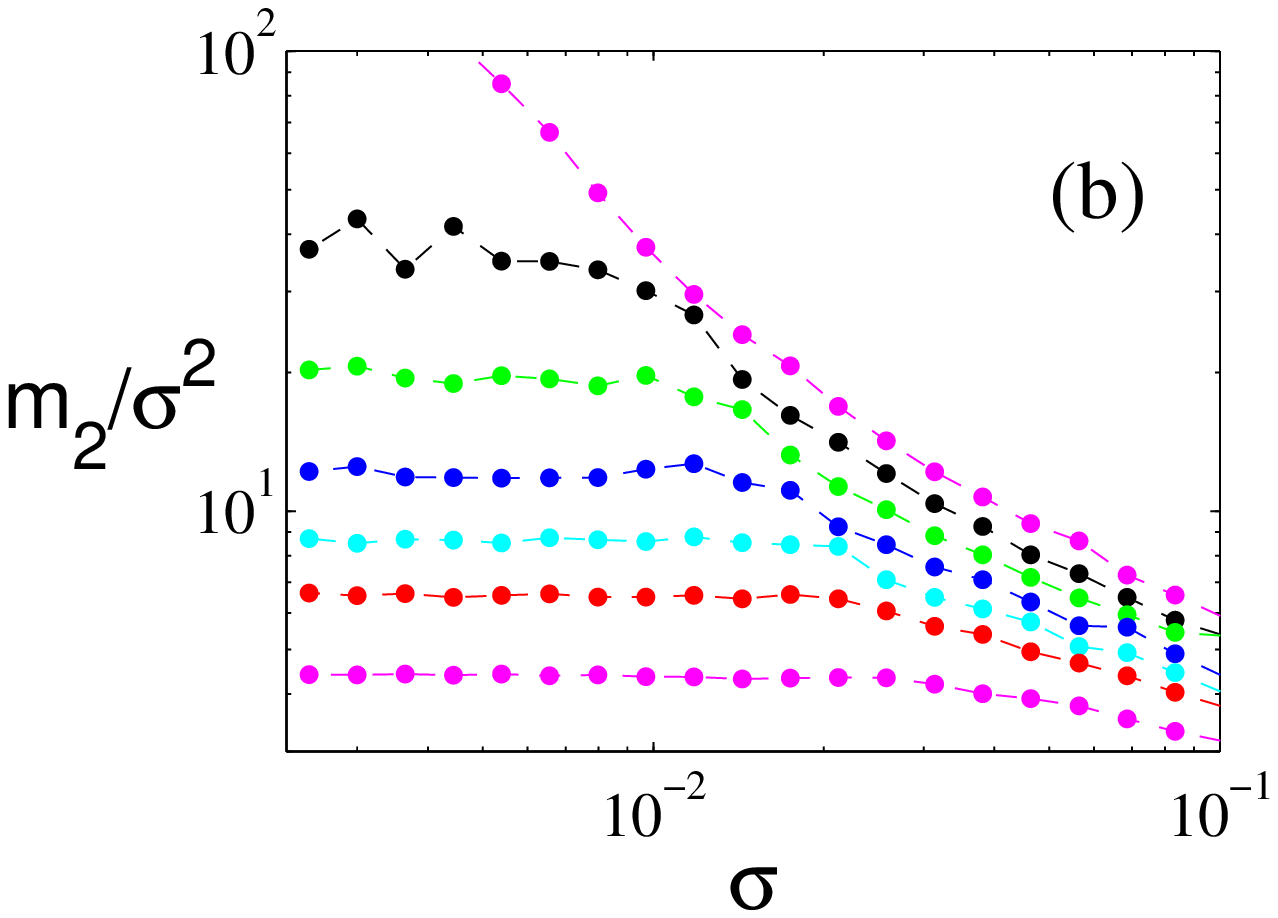,width=.49\textwidth}\hfill
\epsfig{file=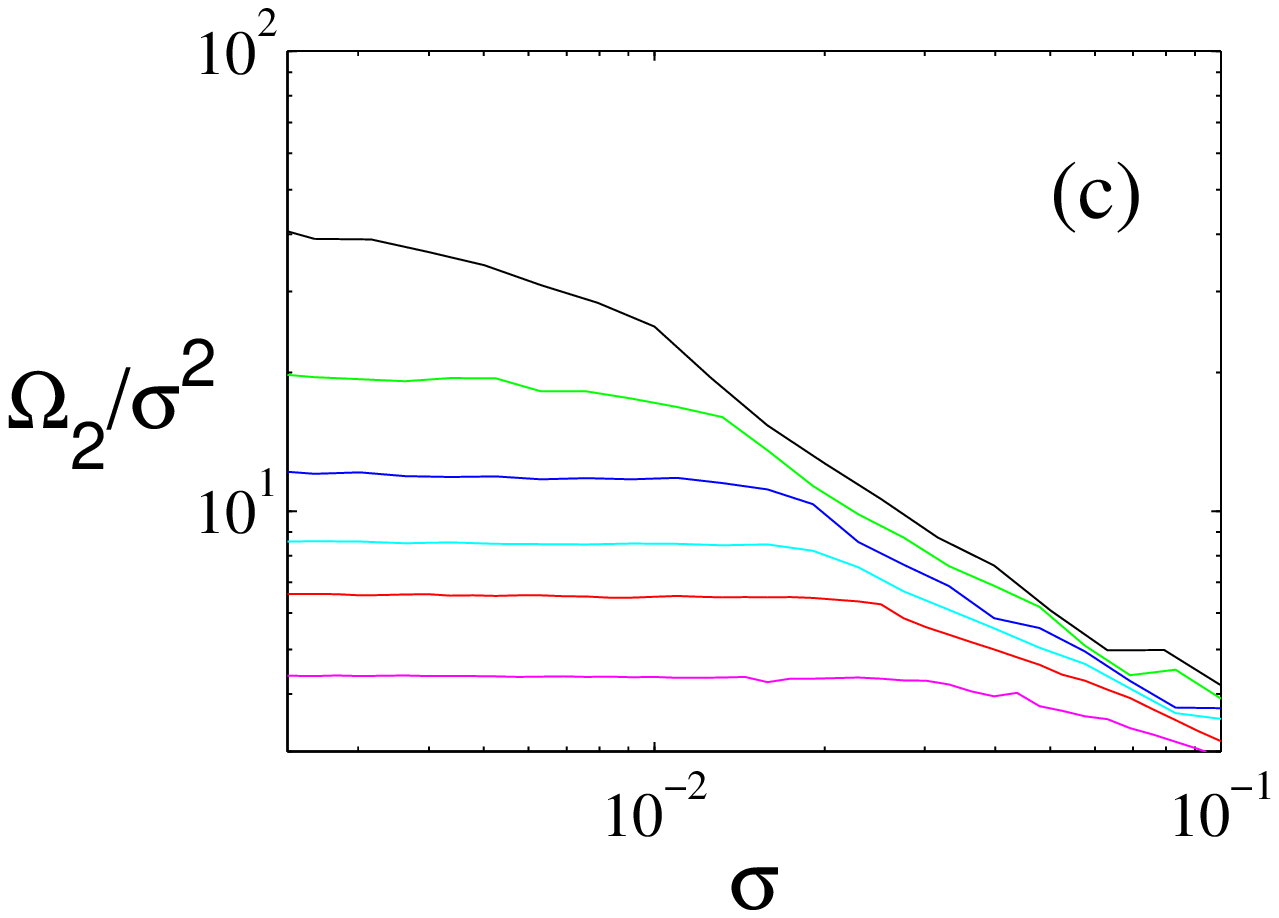,width=.49\textwidth}
\caption{Anomalous scaling in populations of globally-coupled noisy
logistic maps. (a) variance of the individual pdf vs noise strength.
From top to bottom curve: $K=0.3$, 0.31, 0.315, 0.32, 0.33, 0.35, 0.4
(population of about $2^{18}$ maps). For the smallest coupling, 
the second moment tends to a constant value, an indication of 
the existence of clustered solutions.
(b) Ratio between the second moment of the snapshot distribution and
  the noise distribution variance $\sigma^2$
  (from top to bottom $K=0.32,0.34,0.36,0.38,0.4,0.42,0.46$). 
These data have been obtained by direct simulation of the 
Perron-Frobenius operator; 
(c) same as (b) but from simulations of Eq.\ (\ref{eq:log2})
(from top to bottom $K=0.34,0.36,0.38,0.4,0.42,0.46$). For too low coupling
  (e.g. $K=0.32$) the reduced system diverges.}
\label{fig:scalk}
\end{figure}

\end{section}

\begin{section}{Collective dynamics of populations with different kinds of
    microscopic disorder}\label{sec:cfr} 

Parameter mismatch and noise are often viewed as two fundamental ways of
representing microscopic disorder. In the first case, the mismatch within
the population is traced back to the existence of a time-independent
distribution in the properties of single-elements. In the second
case, the difference among individuals is due to some dynamical 
processes that cannot be described at the chosen observation
level and are generically described by noise terms.
The persistence of the main features of the population dynamics after the
introduction of a small amount of individual diversity and/or noise 
is usually
regarded as a proof of the robustness of a phenomenon and, consequently,
of its relevance in the description of real-world systems. 
Here, we propose to use the order parameter expansion approach to
perform a systematic comparison between the effects of these two sources of
disorder. 
The reduced system for noisy maps derived in Sect.~\ref{sec:nope} is
compared to that obtained, under different closure assumptions
\cite{demonte02,demonte03}, for populations with parameter mismatch. 
To this aim, we draw the microscopic disorder according to
the same assigned distribution in both cases.

The equations describing populations with parameter mismatch have the same
form as those with additive noise Eq.\ (\ref{eq:npop}):
\begin{eqnarray} \label{eq:ppop} 
x_j\mapsto (1-K)\,f(x_j)+K\,\nsum{f(x)}+ p_j \quad j=1,\dots N, 
\end{eqnarray}
but the additive term $p_j\in \mathbb R$ is now assigned once and for
all. According to the choice we made for the case of noisy maps, we consider a
uniform disorder distribution  
with mean $p_0=0$, variance $\sigma^2$ and kurtosis $m_4$. The 
parameters $p_j$ will be taken equally spaced on the support of
such a distribution in order to let the bifurcation diagram be insensitive to
the fluctuations related to a resampling.   

To obtain the equations governing the mean field evolution, we apply
again an order parameter expansion by averaging Eq.\ (\ref{eq:npop}) and by
performing the change of variables $\epsilon_j=x_j-X$.
Analogously to what was done in the more general case of a dependence of
$f$ on the distributed parameter \cite{demonte03}, the expansion can be closed
under the assumption that the collective dynamics is coherent. The coherency
condition $||x_j-X||\ll 1$
states that the population dynamics keeps all the individuals within a small
neighborhood of the mean field state.
This assumption allows us to neglect terms of degree larger than one in
$\epsilon_j$, independently of the population size.
\\
The reduced system for parameter mismatch has the form:
\begin{align} \label{eq:popemaps}
\begin{cases}
\rule[-4mm]{0mm}{6mm}
X\mapsto&\: 
f(X)+\frac 1 2 \left[d_x f(X)\right]^2\,\Omega_2\\
\rule[-4mm]{0mm}{6mm}
W\mapsto&\: \sigma^2+(1-K)\,d_x f(X)\,W\\ 
\Omega_2\mapsto& \:\sigma^2+2\,(1-K)\,d_x f(X)\,W+(1-K)^2\\
&\times\left\{\left[d_x f(X)\right]^2\,\Omega_2
+\frac 1 4 \left[d_{xx} f(X)\right]^2 
\left(m_4-\Omega_2^2\right)\right\}.\rule{0mm}{6mm}\end{cases}
\end{align}
Hence, the mean field is coupled to the second moment of the snapshot
distribution $\Omega_2=\nsum{\epsilon_j^2}$, which is, in turn,
influenced by the ``shape'' order parameter 
$W=\nsum{p_j\,\epsilon_j}$ \cite{demonte03}.
As for the reduced system Eq.\ (\ref{eq:log2}), obtained in the case of noise,
the population-level parameters that naturally emerge in the 
expansion as macroscopically relevant are the coupling strength $K$
and the moments, namely the variance $\sigma^2$ and fourth moment
$m_4$, of the parameter distribution.

In the case of logistic maps $f(x)=1-ax^2$ the reduced system to
second order Eq.\ (\ref{eq:popemaps}) is:
\begin{align} \label{eq:logpope}
\begin{cases}
\rule[-4mm]{0mm}{6mm}
X\mapsto&\: 1-a\,X^2-2\,a\,\Omega_2\\
\rule[-4mm]{0mm}{6mm}
W\mapsto&\: \sigma^2-2\,a\,(1-K)\,X\,W\\ 
\rule[-2mm]{0mm}{6mm}
\Omega_2\mapsto& \:\sigma^2-4\,a\,(1-K)\,X\,W
+a^2\,(1-K)^2\,\left[m_4+\left(4\,X^2-\Omega_2\right)\,\Omega_2\,\right]
\end{cases}.
\end{align}
For maximal coupling, this description converges to the same
scalar equation as obtained for noisy populations. In this 
limit, the shape parameter and the population variance become uncoupled 
and coincide. From a 
microscopic point of view, this corresponds to the fact that the snapshot pdf
for $K=1$ in both cases equals the 
distribution of disorder, that is solidly displaced with the mean field
motion (hence, Figs.\ \ref{fig:npdfs} (a-b) provide a microscopic picture
valid for both sources of disorder). The only difference is that in
the case of parameter mismatch the oscillators maintain the order of their
arrangement, 
while in the case of noise they randomly exchange their positions at each time
step. A straightforward consequence of this is that even if the macroscopic
dynamics is indistinguishable for infinite population size, in the case of
noise it depends essentially on the number of elements composing the
population, while this dependence is much weaker for parameter diversity. In
the following, we discuss why this difference exists for every value of the
coupling.

\begin{figure}[h]
\center
\epsfig{file=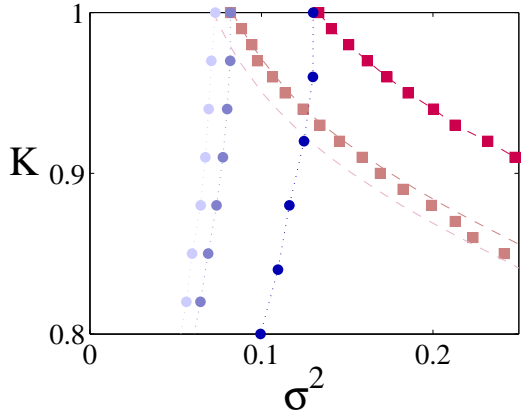,width=.5\textwidth}
\caption{Period-doubling bifurcation lines for the population of noisy maps
  studied in Sec.\ \ref{sec:phenomenology} (full system, dots and reduced
  system to the second degree Eq.\ (\ref{eq:log2}), dotted lines) and for Eq.\
  (\ref{eq:ppop}) (population of $N=64$ logistic maps, squares and reduced system 
  Eq.\ (\ref{eq:logpope}), dashed lines). The bifurcation parameters are the
  variance of the parameter or noise distribution $\sigma^2$ and the coupling
  strength $K$. In the top-right region of the diagram the mean field is
  stationary for both forms of disorder, while for still larger $\sigma ^2$
  the macroscopic dynamics diverges.\label{fig:cfrbif}}   
\end{figure}

When the coupling is less than maximal, 
on the other hand, we expect to observe
differences in the macroscopic regimes induced by parameter mismatch and
noise. These differences can be visualized by plotting in the plane
$(\sigma^2,K)$ the period doubling bifurcation lines for the mean field motion
(Fig.\ \ref{fig:cfrbif}).
While for weaker couplings a low intensity of noise is sufficient to
drive the system out of the chaotic region, the opposite happens if  
the microscopic disorder is due to a diversity in the parameters. Indeed, in
this case the region where the mean field displays chaotic behaviour becomes
larger for smaller $K$.
This property of the macroscopic dynamics is captured by the reduced system
Eq.\ (\ref{eq:popemaps}), whose bifurcation lines provide a good quantitative
approximation of those determined by numerically simulating the population.  

As mentioned above, another important difference between the two sources of
microscopic disorder 
is the dependence of the collective dynamics on the population size. This is
captured by the difference in the condition of validity of the reduced
systems that describe the macroscopic dynamics. As 
pointed out in Sec.\ \ref{sec:nope}, the derivation of Eq.\
(\ref{eq:redpol}) is based on the assumption that the population is
infinitely large, while the mean field of finite populations undergoes
fluctuations that scale as $1/\sqrt{N}$. 
On the other hand, Eq.\ (\ref{eq:redpol}) has been derived without making
any assumption on the system size, but rather under the hypothesis that the
collective regimes are coherent. Hence, in the case of parameter mismatch the
system size is expected to play a minor role for the qualitative features of
the macroscopic dynamics. 
The weak dependence on the system size of the mean field behaviour for
populations with parameter mismatch is
confirmed by the numerical simulations and can be understood by looking at the
characteristics of the snapshot distribution. 
Figure\ \ref{fig:pdfscy} displays
the instantaneous pdfs for the populations with parameter 
mismatch and additive
noise in regimes where the mean field has a period-two attractor.  
If the disorder is due to parameter mismatch, the shape of the instantaneous
pdf is little affected by the number of points that identify its profile,
along which the oscillators maintain their ordered configuration.  
If noise is present, on the other hand, reducing the population size alters the
shape of the instantaneous pdf at every time step, broadening it
with respect to the infinite-size limit distribution, and, hence,
affecting the mean field dynamics in an unpredictable way.

\begin{figure}[h]
\center
\epsfig{file=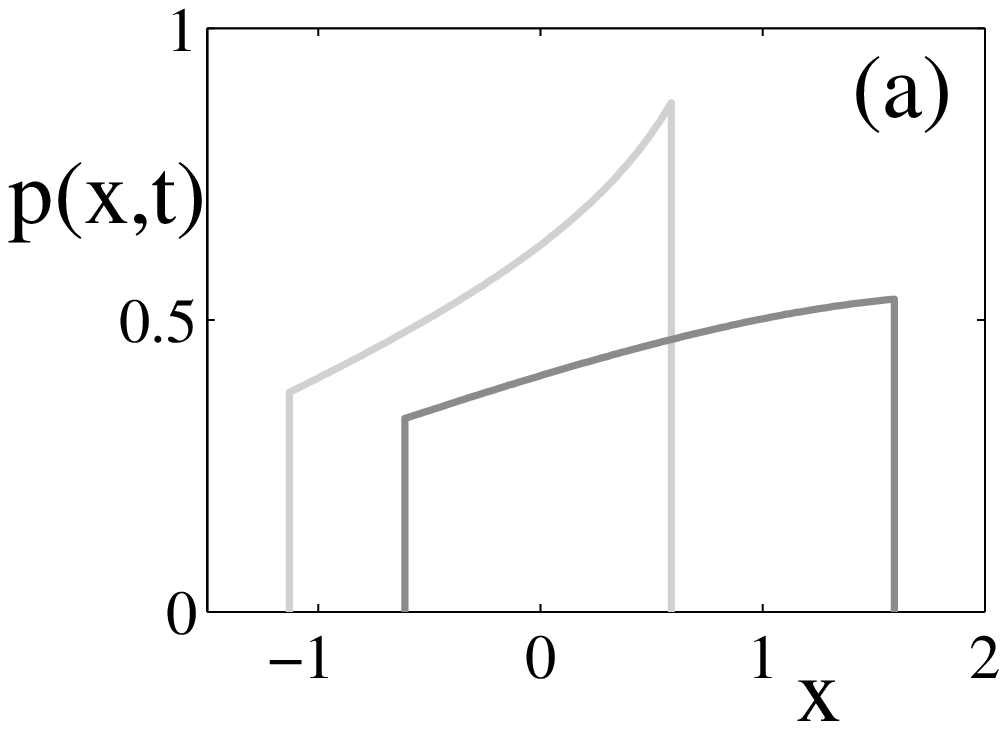,width=.3\textwidth}
\epsfig{file=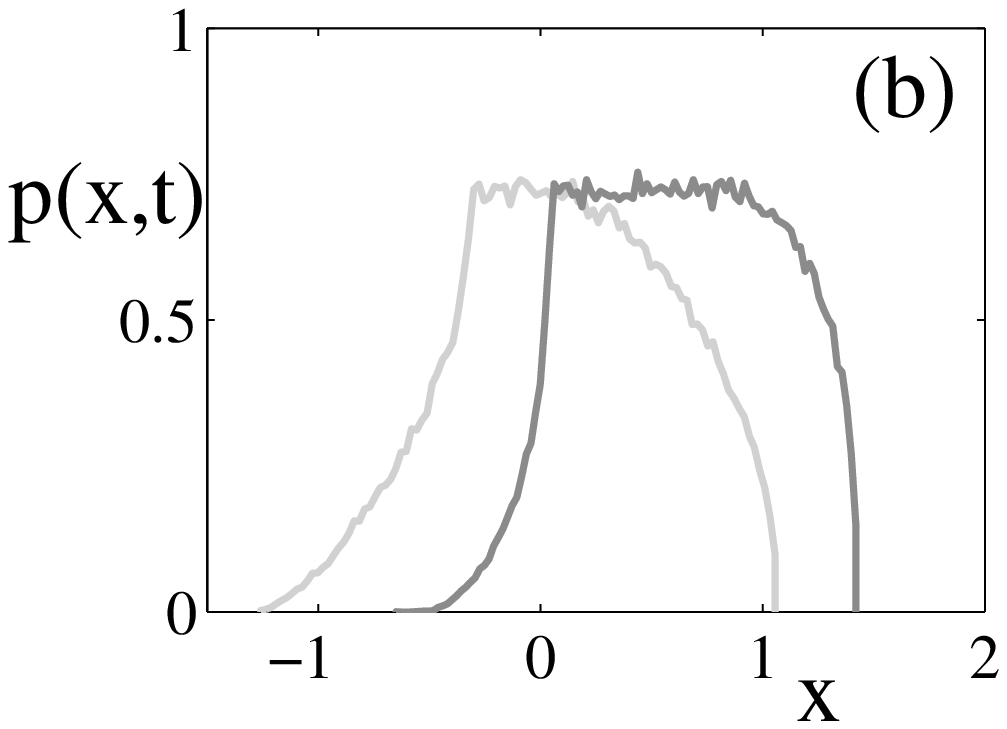,width=.3\textwidth}

\epsfig{file=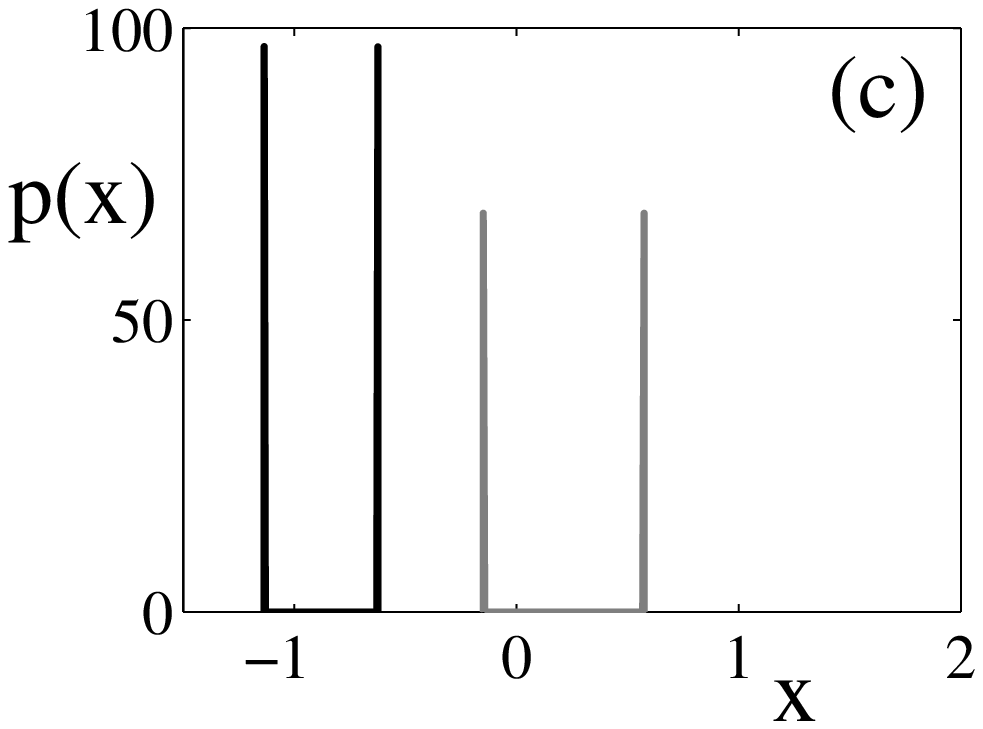,width=.3\textwidth}
\epsfig{file=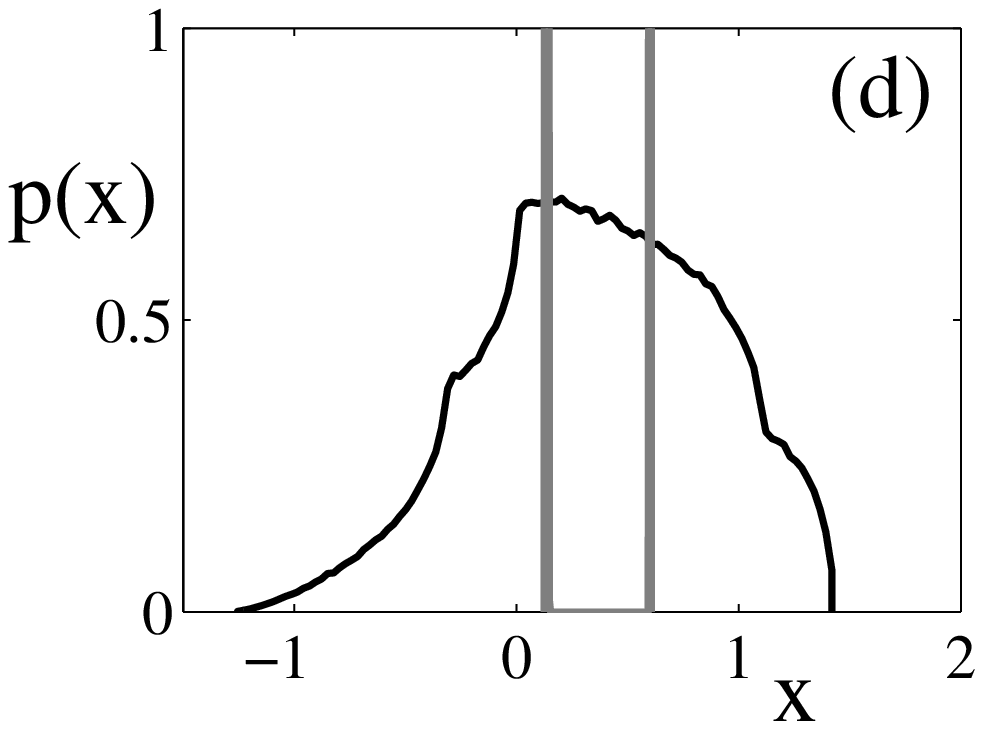,width=.3\textwidth}
\caption{Snapshot pdfs (at even and odd times) (a-b) and mean field and
  individual pdfs (c-d) for a macroscopic
  period-two   regime in populations of 
  logistic maps with (a-c) parameter mismatch ($\sigma^2=0.6$, $K=0.9$) and
  (b-d) additive noise ($\sigma^2=0.4$, $K=0.7$). 
\label{fig:pdfscy}}
\end{figure}

Another conclusion that can be drawn by looking at Figs.\
\ref{fig:pdfscy} (a) and (b) is that the snapshot distribution 
reflects the source of
microscopic diversity that underlies collective behaviour. Thus, given a
certain mean-field dynamics one 
should be able, in principle, to identify the character of the
microscopic disorder by looking at the statistical properties of the
instantaneous distribution.
This is even more evident if we compare the individual pdfs,
obtained from the time series of an individual map within the 
populations with different nature of microscopic disorder. Figures\
\ref{fig:pdfscy} (c) and (d) illustrate this in the case when the two
populations possess qualitatively similar macroscopic regimes, that is
they are both periodic of period two. Correspondingly, the
mean field pdf consists in two Dirac delta functions centered on the values
assumed
by the cycle. Although the two populations cannot be distinguished at
a macroscopic level, the difference is evident if one considers the
individual pdfs.
In the case of parameter mismatch, the probability distribution of an
individual time series is identical to the mean field pdf, but centered
on shifted values. This is a consequence of the 
fact that one oscillator in the population maintains its
position relative to the population average.
When the disorder is due to noise, on the other hand, the individual pdf is
broad, since the trajectory of one individual within the population is
blurred by the stochastic term.
Having access to a macroscopic-level and a microscopic-level
observable, such as the time series of the average and of one element
within the population, these differences can be used to identify the kind of
disorder that is the main determinant of the
emergent population dynamics.  

\end{section}

\begin{section}{Discussion}\label{sec:discuss}

In this paper, we have described the transition from full synchronization to
disorder-induced collective dynamics in large populations of globally coupled
maps, a prototype for the study of systems with many degrees of freedom. 
The influence of microscopic variability on the macroscopically accessible
degrees of freedom is particularly relevant in modeling biological
populations and in the interpretation of experimental results.

Here we have mostly treated the case of additive noise, 
and focused on its influence on the collective dynamics, a 
viewpoint complementary to that adopted by Teramae and Kuramoto
\cite{teramae01}. 
We argued that noise induces changes in the macroscopic dynamics which 
are probably at the origin of the anomalous scaling properties 
reported by these authors. 

We have presented the systematic derivation of finite-dimensional
reduced systems which were showed to account quantitatively
and with increasing accuracy for the collective dynamics of the full system.
The reduced systems link the dynamics of the mean-field to 
microscopic properties of the population,
such as the single-element dynamics parameters and the moments of the noise
distribution. Remarkably, the reduced systems were shown to
exhibit anomalous scaling themselves. We believe this is because
both the full and the reduced system share the folded, fractal 
phase-space structure at the origin of the macroscopic bifurcations leading to
anomalous scaling. This conjecture should be addressed in the future,
in an effort to link the scaling properties of the collective dynamics
attractors to the anomalous scaling of the population distribution.

Finally, we addressed the effects on the collective dynamics of
strongly coupled maps of different sources of microscopic disorder. The
results obtained on the noise-induced mean-field bifurcations have
been compared to those obtained when the noise is ``quenched'', that is when
there is a parameter mismatch within the population. 
Even though in both cases the macroscopic dynamics
is qualitatively modified as the disorder is increased, these changes differ in
general depending on the origin of the microscopic disorder
and we argued that this can be used in the interpretation of experimental
data, even when only macroscopic observables are accessible.

The analytical approach presented in this paper allows us to derive in a
systematic manner macroscopic equations ruling the effective dynamics of
populations of globally and strongly coupled dynamical system. Although here
we only presented results relative to maps, we envisage
that a similar approach might be applied to continuous-time systems.

{\bf Acknowledgments}

S. De Monte acknowledges support from the ESF program REACTOR and from the
Danish Natural Science Foundation.

\end{section}

\appendix
\begin{section}
{Order parameter expansion for noisy maps}\label{app:ope} 

From Eq.\ (\ref{eq:omega}), one can see that, if the map is a
polynomial of order $P$ and the expansion is truncated
to the $n$-th degree in $(1-K)$, then the mean-field is coupled to the order
parameters from 
the second to the $P$-th. The iterates of such $n-1$ variables contain order
parameters from the second up to the $nP$-th degree. The dimensionality of the
macroscopic map is hence equal to $nP$, while the remaining infinite degrees
of freedom, of larger degree, are constantly equal to the moments of the noise
distribution.  

We now demonstrate that such a
$nP$-dimensional truncation can be simplified further and reduced to 
a system of $n$ equations independently of the degree of nonlinearity of the
single-element dynamics. 
In order to show this, let us define the vector ${\bf
\Gamma}$ containing the terms that multiply $(1-K)^i$ ($i<nP$) in the
following way: 
\begin{eqnarray*}
\Gamma_1=&&0\\
\Gamma_i=&&\left\langle \left[\sum_{p=1}^P\frac{1}{p!}\ma{D}^pf(X)
\left(\epsilon^p-\Omega_p\right)\right]^i\right\rangle
\hspace{5mm} i=2,\dots, n\,P. \nonumber
\end{eqnarray*}
Let us call ${\bf \Omega}=\left\{\Omega_1, \dots \Omega_{nP}\right\}$ the
vector collecting the first $n\,P$ order 
parameters and ${\bf m}$ the vector
composed by the corresponding moments of 
the noise distribution. We will show that only the first $n-1$ order
parameters are 
independent variables in the macroscopic dynamics, while the other elements
of the vector $\Omega$ can be expressed as linear combinations of appropriate
functions of the first ones.  

The truncation to $n$-th degree of Eq.\ (\ref{eq:omega}) can be written as:
\begin{eqnarray}\label{eq:appgamma}
{\bf \Omega}-{\bf m}\mapsto\ma{C}\:{\bf \Gamma},
\end{eqnarray} 
where $\ma{C}$ is an $nP \times nP$ matrix whose entries are:
\begin{eqnarray*}
\begin{cases}
\vspace{1mm}
\ma{C}_{q,i}=\binom{q}{i}\left(1-K\right)^i\:m_{q-i}\hspace{10mm}
&q=2,\dots ,n\,P\qquad i<\min \{ q,n \}\\
\vspace{1mm}
\ma{C}_{q,q}=(1-K)^q  &2\le q\le n\\
\ma{C}_{q,i}=0 &\text{otherwise}.
\end{cases}
\end{eqnarray*}

The truncation of the equations for the order parameters have a
number of constants of motion equal to the dimension of $\ker
(\ma{C})$. Hence, it can be reduced by means of a linear transformation to a
system whose dimension equals the rank of $\ma{C}$.
It is easy to see that $\ma{C}$ has rank smaller or equal to $n-1$, since the
first column and all
the columns with index $i\in\{n+1, \dots n\,P\}$ have only null entries.
Moreover, the rank of $\ma{C}$ is exactly $n-1$ since all the $n-1$
diagonal elements with index $i\in\{2, \dots n\}$ 
are nonzero as long as $K<1$.

Hence, the $n-1$ order parameters of second to $n$-th degree are independent
variables coupled to the mean field, so that the reduced system is of
dimension $n$. The order parameters of $n+1$-th to $n\;P$-th degree are
dependent variables and their dynamics is slaved to that of the reduced
system. 
Such slaved degrees of freedom can be obtained as linear combinations of the
independent order parameters by projecting Eq.\ (\ref{eq:appgamma}) onto the
null space, and solving a set of $(n-1)\,P$ equations in the same number of
variables. 

\end{section}

\bibliographystyle{plain}
\bibliography{noise,synch}

\end{document}